\documentclass[11pt]{article}

\usepackage{fullpage,url,graphicx,amssymb,amsmath,cite}
 \usepackage{mdframed}

\def\bbR{\mathrm{R}}

\def\dx{\mathrm{d}x}
\def\dy{\mathrm{d}y}

\def\erf{\mathrm{erf}}
\def\erfc{\mathrm{erfc}}
\def\simiid{\sim_\mathrm{iid}}
\def\floor#1{{\lfloor {#1} \rfloor}}
\def\bbN{\mathbb{N}}
\usepackage{ifpdf}
\ifpdf
\fi

\title{A series of maximum entropy upper bounds of the differential entropy}

\author{Frank Nielsen
\thanks{Frank Nielsen is with \'Ecole Polytechnique (Palaiseau, France)
and Sony Computer Science Laboratories Inc. (Tokyo, Japan). Email: {\tt Frank.Nielsen@acm.org}}
\and Richard Nock\thanks{Richard Nock is with Data61, the Australian National University, and the University of Sydney, Australia. Email: {\tt Richard.Nock@data61.csiro.au}}}
\date{}

\begin{document}

\maketitle 

\begin{abstract}
We present a series of closed-form maximum entropy upper bounds for the differential entropy of a continuous univariate random variable and study the properties of that series. 
We then show how to use those generic bounds for upper bounding the differential entropy of Gaussian mixture models.
This requires to calculate the raw moments and raw absolute moments of Gaussian mixtures in closed-form that may also be handy in statistical machine learning and information theory.
We report on our experiments and discuss on the tightness of those bounds.
\end{abstract}


\def\ceil#1{ \lceil #1 \rceil }
\def\erf{\mathrm{erf}}
\def\erfc{\mathrm{erfc}}
\def\AMEF{\mathrm{AMEF}}
\def\MEUB{\mathrm{MEUB}}
\def\floor#1{{\lfloor {#1} \rfloor}}
\def\bbN{\mathbb{N}}
\def\dx{\mathrm{d}x}
\def\dy{\mathrm{d}y}
\def\inner#1#2{\langle #1,#2\rangle}
\def\overbar{\overline}
\def\calX{\mathcal{X}}
\def\bbR{\mathbb{R}}

\def\MaxEnt{\mathrm{MaxEnt}}
\newtheorem{example}{Example}
\newtheorem{remark}{Remark}

\newmdtheoremenv{theorem}{Theorem}
\newmdtheoremenv{lemma}{Lemma}
\newmdtheoremenv{corollary}{Corollary}

\def\st{\ : \ }

\section{Introduction}
Shannon's {\em differential entropy}~\cite{ct-2012} $H(X)$ of a {\em continuous random variable} $X$ 
following a probability density function $p(x)$ (denoted by $X\sim p(x)$) on the support 
$\calX=\{x\in \bbR \ :\ p(x)>0\}$ quantifies the amount of uncertainty~\cite{ct-2012} of $X$ by the following celebrated formula:
 
\begin{equation}\label{eq:H}
H(X) = \int_\calX p(x)\log\frac{1}{p(x)}\dx = -\int_\calX p(x)\log p(x)\dx .
\end{equation}
When the logarithm is expressed in basis $2$, the entropy is measured in {\it bits}. 
When using the natural logarithm (basis $e$), the entropy is measured in {\it nats}.
The entropy functional $H(\cdot)$ is {\em concave}~\cite{ct-2012}, 
may be {\em negative}\footnote{For example, when  $X\sim N(\mu,\sigma)$ is a Gaussian distribution of mean $\mu$ and standard deviation $\sigma>0$, then $H(X)=\frac{1}{2}\log(2\pi e\sigma^2)$, and is therefore negative when $\sigma<\frac{1}{\sqrt{2\pi e}}$.}, and may be  {\em infinite}\footnote{For example, consider $X\sim p(x)$ with $p(x)=\frac{\log(2)}{x\log^2 x}$ for $x>2$ (with support $\calX=(2,\infty)$). Then $H(X)=+\infty$. 
This result is to contrast with the fact that the discrete entropy on a finite alphabet $\calX$ is bounded by $\log |\calX|$.} when the integral of Eq.~\ref{eq:H} diverges.

Although closed-form formula for the differential entropy are available for many common statistical distributions (see the devoted book~\cite{H-2013} and~\cite{H-EF-2010}), the {\em differential entropy of mixtures} usually does not admit  closed-form expressions~\cite{KLnotanalytic-2004,H-GMM2-2008} because the log term in Eq.~\ref{eq:H}   transforms into an untractable {\em log-sum term} when dealing with mixture densities.
Let us denote by $m(x)=\sum_{c=1}^k w_cp_c(x)$ the density of a mixture\footnote{Beware that the mixture random variable $M\not =\sum_i w_iX_i$. The  probability density of a weighted sum of random variables is obtained by convolution of the densities.} $M\sim m(x)$ with $k$ components $X_c\sim p_c(x)$, where $w\in\Delta_k$ denotes the $k$-dimensional open probability simplex.
That is, a mixture is a {\em convex combination} of component distributions $p_1(x),\ldots, p_k(x)$. 
We shall consider mixtures of Gaussians  with component probability density functions $X_i\sim N(\mu_i,\sigma_i)$ such that:  
$$
p_i(x)=p(x;\mu_i,\sigma_i)=\frac{1}{\sqrt{2\pi}\sigma_i}\exp\left(-\frac{(x-\mu_i)^2}{2\sigma_i^2}\right),
$$
 where $\mu_i=E[X_i]\in\bbR$ and $\sigma_i=\sqrt{E[(X_i-\mu_i)^2]}>0$ denote the mean parameter and the standard deviation of $X_i$, respectively.

Statistical mixtures  allow flexible fine modeling of {\em arbitrary} smooth densities: They are provably universal smooth density estimators.
The most common mixtures are the Gaussian Mixture Models (GMMs) that are frequenty met in applications.
To tackle the differential entropy of continuous mixtures, various {\em approximation techniques} have been designed  (see~\cite{KL-M-2016} and references therein for a state-of-the-art).
In practice, to estimate $H(X)$ with $X\sim p(x)$, one uses the following {\em Monte-Carlo (MC) stochastic integration}:
\begin{equation}\label{eq:Hsto}
\hat H_s(X)=-\frac{1}{s}\sum_{i=1}^s \log p(x_i),
\end{equation}
where $\{x_1, \ldots, x_s\}$ is an independent and identically distributed (iid) set of variates sampled from $X\sim p(x)$.
This MC estimator $\hat H_s(X)$ is consistent  (ie., $\lim_{s\rightarrow\infty} \hat H_s(X)=H(X)$, convergence in probability).
Moshksar  and  Khandani~\cite{H-GMMiso-2016} recently considered the special case of isotropic spherical Gaussian Mixture Models (ie., GMMs with identical standard deviation),
 and used Taylor expansions to arbitrarily finely approximate the differential entropy of those isotropic GMMs. 
Interestingly, they mentioned in their paper~\cite{H-GMMiso-2016} the so-called {\em Maximum Entropy Upper Bound} (MEUB) that relies on the fact that the continuous distribution with prescribed variance maximizing the entropy is the Gaussian distribution of same variance. 
Since the entropy of a univariate Gaussian $N(\mu,\sigma)$ is $\frac{1}{2}\log(2\pi e\sigma^2)$, we end up with the following maximum entropy upper bound for an arbitrary random variable $X$:
\begin{equation}\label{eq:MEUB}
H(X)\leq \frac{1}{2}\log\left(2\pi e V[X]\right), 
\end{equation}
where $V[X]=E[(X-E[X])^2]=E[X^2]-E[X]^2$ denotes the variance of $X$.
Since the variance $V[X]$ of an arbitrary Gaussian mixture can be easily calculated in closed-form~\cite{KL-M-2016}:
$$
V[X]=\sum_{c=1}^k w_c\left((\mu_c-\bar\mu)^2+\sigma_c^2)\right),
$$ 
with $\bar\mu=\sum_{c=1}^k w_c\mu_c=E[X]$), Eq.~\ref{eq:MEUB} yields the {\em Gaussian MaxEnt  Upper Bound}:
\begin{equation}\label{eq:megub}
H(M)\leq  \frac{1}{2}\log\left(2\pi e \sum_{i=1}^k w_i((\mu_i-\bar\mu)^2+\sigma_i^2))\right).
\end{equation}

In this work, we propose to further use the maximum entropy upper bound principle to derive an {\em infinite series} of MaxEnt upper bounds.
Although our bounds will be instantiated for GMMs, they apply more broadly to univariate continous mixtures.
For example, our MaxEnt upper bounds  also hold for mixtures of exponential families~\cite{MEF-1980}  that generalize the GMMs (and have always guaranteed finite entropy).
For GMMs, we shall show that the Gaussian MaxEnt Upper Bound is not necessarily the best MaxEnt upper bound in closed form, and report instead a series of upper bounds.

The paper is organized as follows:
Section~\ref{sec:maxent} introduces the general principle for building  Maximum Entropy (MaxEnt) Upper Bounds.
It is followed by Section~\ref{sec:rgm} that construct a series of MaxEnt upper bounds  derived from a special family of MaxEnt distributions that we termed Absolute
Monomial Exponential Families.
Those generic bounds are instantiated for Gaussian Mixture Models (GMMs) in Section~\ref{sec:mm}.
Section~\ref{sec:exp} report on our experiments and discusses the tightness of the bounds.
Finally, Section~\ref{sec:concl} wrap ups the results and conclude the work.
Besides, an appendix provides the detailed calculation of the raw absolute  moment of a non-centered normal distribution that is used in Section~\ref{sec:mm} to get closed-form MaxEnt upper bounds for GMMs.

\section{Maximum entropy upper bounds on the differential entropy\label{sec:maxent}}

The MaxEnt distribution principle was investigated by Jaynes~\cite{maxent1-1957,maxent2-1957}  to infer a distribution given several ``moment constraints.'' 
MaxEnt asks to solve the following constrained optimization problem:
\begin{equation}\label{eq:maxent}
\max_p H(p) \st E[t_i(X)]=\eta_i,\quad i\in [D]=\{1,\ldots,D\}.
\end{equation}
When an iid sample set $\{x_1, \ldots, x_s\}$ is given, we may choose, for example, the raw geometric {\em sample moments} 
$\eta_i=\frac{1}{s} \sum_{j=1}^s x_j^i$ for setting up the constraint $E[X^i]=\eta_i$ (ie., taking $t_i(X)=X^i$ in Eq.~\ref{eq:maxent}).
The distribution $p(x)$ maximizing the entropy under those moment constraints is unique and termed the {\em MaxEnt distribution}.
The constrained optimization of Eq.~\ref{eq:maxent} is solved by means of Lagrangian multipliers~\cite{maxent-1992,ct-2012}.
It is well-known~\cite{maxent-1992,ICA-2001} that the MaxEnt distribution $p(x)$ belongs to  a {\em parametric family} of distributions called an {\em exponential family}~\cite{EF-1986}.
An exponential family (EF) admits the following {\em canonical probability density function}:

\begin{equation}\label{eq:ef}
p(x;\theta) = \exp\left(\inner{\theta}{t(x)}-F(\theta)\right),
\end{equation}
where $\inner{a}{b}=a^\top b$ denotes the scalar product, and $\theta\in\Theta$ the natural parameter vector belonging to the natural parameter space $\Theta\subset \bbR^D$.
For MaxEnt distributions, the exponential family $\{p(x;\theta)\st \theta\in\Theta\}$ is generated by the sufficient statistics $t_i(x)$'s.
The natural parameter of the MaxEnt distribution is given by the Lagrangian multipliers~\cite{maxent-1992,ct-2012}.
It follows that we have $E_X[t(X)]=\eta$ for some unique random variable $X\sim p(x;\theta)$ with $\theta\in\Theta=\{ \theta \st \int \exp(\theta^\top t(x))\dx <\infty \}$, see~\cite{EF-1986}.
The function $F(\theta)=\log\int p(x;\theta)\dx$ is called the log-normalizer~\cite{EF-1986} since it allows to normalize the density to a probability: $\int_\calX p(x;\theta)\dx=1$. In statistical physics,   the partition function $Z(\theta)=\exp(F(\theta))$ is rather used to normalize the distributions.

The key observation is to notice that by construction,  {\em any other distribution} with density $p'(x)$ different from the MaxEnt distribution $p(x)$ and satisfying all the $D$ moment constraints $E[t_i(X)]=\eta_i$ will have necessarily smaller entropy: 
$H(p'(x))\leq H(p(x))$ with $p(x)=p(x;\theta)$.
However, depending on the choosing sufficient statistics $t_i$'s, neither $\theta$ nor $F(\theta)$ may be available in closed-forms, and thus need to be approximated numerically~\cite{maxent-1992}.

In the remainder, we upper bound the differential entropy of a continuous random variable (eg., finite mixtures) by building a {\em collection} of upper bounds derived from  MaxEnt distributions 
which admit closed-form  expressions for their differential entropy.
Those bounds proves handy in practice for GMMs since the differential entropy of a GMM is not available in closed-form~\cite{H-GMMiso-2016,KLnotanalytic-2004}.
Besides, we report closed-form formula for calculating the arbitrary raw absolute moments of a univariate Gaussian Mixture Model that may prove useful in other areas of statistical machine learning and information theory.

\section{MaxEnt upper bounds from raw absolute moment constraints: The Absolute Monomial Exponential MaxEnt distributions\label{sec:rgm}}
Consider the univariate uni-order ($D=1$) family of  {\em Absolute Monomial Exponential Family} (AMEF) induced by the absolute value of a {\em monomial}  of degree $l\in\bbN$ defined over the full support $\calX=\bbR$:

\begin{equation}\label{eq:amefpdf}
p_l(x;\theta)  =  \exp\left(\theta |x^l|-F_l(\theta)\right),\quad x\in\bbR
\end{equation}
for $\theta<0$.
The natural parameter space is $\Theta=(-\infty,0)$).
The log-normalizer\footnote{This integral can be computed using any Computer Algebra Systems (CASs) like {\tt Maxima} that can be downloaded at \url{http://maxima.sourceforge.net/}. CASs implement the semi-algorithm of Risch~\cite{Risch-1970} for symbolic integration, and can therefore determine whether the integral admits a closed-form or not in terms of elementary functions. See~\cite{Bronstein-2013,Davenport-2016} for recent developments. Some CAS code snippets are provided in the appendix.} is:
\begin{eqnarray}
F_l(\theta) = \log 2+\log \Gamma\left(\frac{1}{l}\right)-\log l - \frac{1}{l}\log (-\theta),\label{eq:Fl}
\end{eqnarray}
where the Gamma function $\Gamma(u)=\int_0^\infty x^{u-1}\exp(-x)\dx$ generalizes the factorial ($\Gamma(n)=(n-1)!$ for $n\in\bbN$).
The Gamma function  can be approximated finely in a few constant operations.
In fact, even better, it is the function $\log\Gamma(u)$ that can be calculated quickly (see the numerical receipe in~\cite{nr-2007}), so that the log-normalizer of Eq.~\ref{eq:Fl} can be calculated fast for any $l\in\bbN$ and $\theta<0$.
Note that those AMEF distributions are {\em unimodal} distributions with the unique mode located at $x=0$.
Since $p_l(x;\theta)=p_l(-x;\theta)$, the mean $E_{p_l(x;\theta)}[X]$ of an AMEF is always zero.

Now, the key element is to notice that the differential entropy $H_l(\theta)=H(p_l(x;\theta))$ of an AMEF admits the following closed-form formula:
\begin{eqnarray}
H_l(\theta)&=&\log 2+\log \Gamma\left(\frac{1}{l}\right)-\log l + \frac{1}{l}(1-\log (-\theta)),\nonumber\\
&=& a_l - \frac{1}{l}\log (-\theta), \label{eq:htheta}
\end{eqnarray}
where $a_l=\log 2+\log \Gamma\left(\frac{1}{l}\right)-\log l + \frac{1}{l}$ is a constant independent of $\theta$.
The entropy can be  expressed {\em equivalently} using the Legendre convex conjugate~\cite{H-EF-2010} $F^*(\eta)$   as:
\begin{eqnarray}
H_l(\eta)=-F_l^*(\eta)= F_l(\theta)-\theta F_l'(\theta),
\end{eqnarray}
with $F_l'(\theta)=-\frac{1}{l\theta}=\eta$ and $\theta=-\frac{1}{l\eta}$.
Therefore the entropy formula expressed using the $\eta$-parameter is:
\begin{eqnarray}
H_l(\eta)&=& \log 2 +\log \Gamma\left(\frac{1}{l}\right)-\log l + \frac{1}{l}(1+\log l+\log\eta),\nonumber\\
&=& b_l + \frac{1}{l} \log\eta,  \label{eq:heta}
\end{eqnarray}
with $b_l=\log \frac{2\Gamma(\frac{1}{l})(el)^{\frac{1}{l}}}{l}$ a constant at prescribed $l$, independent of $\eta$.
We readily check Young-Fenchel equality: $F_l(\theta)+F^*_l(\eta)-\theta\eta=0$ since $\theta\eta=-\frac{1}{l}$, and observe that $H_l(\theta)=-F^*_l(\eta)=F_l(\theta)+\frac{1}{l}$.
Let $H=\{\eta(\theta)\st \theta\in\Theta\}$ denote the {\em expectation parameter space}.
For AMEFs, the dual natural/expectation parameter spaces are thus $\Theta=(-\infty,0)$ and $H=(0,+\infty)$.
To avoid confusion, let us denote by $H_l^\theta(\cdot)$ and by $H_l^\eta(\cdot)$ the entropy formula of Eq.~\ref{eq:htheta} and Eq.\ref{eq:heta} with respect to the natural and expectation parameters, respectively.

To upper bound the entropy of {\em any arbitrary univariate continuous random variable} $X$ (let it be a mixture or not), 
we simply calculate the  $l$-th raw absolute geometric moment $A_l(X)=E_{X}[|X|^l]$, 
and deduce the following MaxEnt entropy Upper Bound (MEUB) $U_l$: 
$$
H(X)\leq H_l^\eta\left(E_{X}\left[|X|^l\right]\right)
$$

We thus obtain an infinite countable series of MaxEnt  Upper Bounds (MEUBs) that we summarize in the following theorem:

\begin{theorem}[AMEF MaxEnt Upper Bounds]\label{theo1}
Let $X$ be a continuous random variable with support $\calX=(-\infty,\infty)$. 
Then the differential entropy $H(X)$ of $X$ is upper bounded by the following series of MaxEnt upper bounds:
\begin{equation}\label{eq:ul}
 H(X)\leq U_l(X) = b_l+\frac{1}{l}\log E_X[|X|^l], \quad\forall l\in\bbN,
\end{equation}
where $b_l=\log 2 +\log \Gamma\left(\frac{1}{l}\right)-\log l + \frac{1}{l}(1+\log l)$.
\end{theorem}

Note that for even integer $l$, $A_l(X)=E_{X}[|X|^l]=E_{X}[X^l]$.
That is, the absolute geometric moments coincide with the geometric moments for even integer $l$.

Let us give two well-known MaxEnt distributions that are AMEF  MaxEnt distributions  in disguise, with their corresponding differential entropies:
\begin{itemize}
\item Consider $l=2$.  
Since  $\Gamma(\frac{3}{2})=\frac{\sqrt{\pi}}{2}$, we get $F_2(\theta)=\frac{1}{2}\log\frac{\pi}{-\theta}$.
Thus $p_2(x;\theta)= \sqrt{\frac{-\theta}{\pi}}\exp(\theta x^2)$.
By setting $\theta=-\frac{1}{2\sigma^2}$, we get the usual canonical  {\em  standard Gaussian density}: 
$\frac{1}{\sqrt{2\pi}\sigma}\exp(-\frac{x^2}{2\sigma^2})$.
Since $\theta=-\frac{1}{2\sigma^2}$, we recover the usual entropy of a Gaussian: $H({p}_2(x;\theta)) =\frac{1}{2}\log 2\pi e\sigma^2$.
It follows that:
$$
H({p}_2(x;\theta)) =\sqrt{\frac{-\theta}{\pi}}\frac{\sqrt{\pi}}{2\sqrt{-\theta}}+\log\sqrt{\frac{\pi}{-\theta}} =\frac{1}{2}\log\frac{\pi}{-\theta}e.
$$
Since $\theta=-\frac{1}{2\sigma^2}$, and considering the location-scale family, we get back the usual entropy expression of a Gaussian $X\sim N(\mu,\sigma)$ : $H({p}_2(x;\theta)) =\frac{1}{2}\log 2\pi e\sigma^2$.

\item Consider $l=1$. The MaxEnt distribution is the  {\em standard Laplacian distribution}~\cite{ICA-2001}
 with density written canonically as $p(x;\theta)= \exp\left(\theta |x|-\log(-\frac{2}{\theta})\right)$ with $F_1(\theta)=\log(-\frac{2}{\theta})$.
The differential entropy can be expressed in {\em either} the natural or expectation coordinate system as
$H(p(x;\theta)) =  1+\log\left(\frac{2}{-\theta}\right)$ or $H(p(x;\eta)) =  1+\log(2\eta)$ with $\eta=F_1'(\theta)=-\frac{1}{\theta}$, respectively.
\end{itemize}

In general, the differential entropy of an AMEF distribution of degree $l$ is negative when:
$$
\eta<\frac{l^l}{le (2\Gamma(\frac{1}{l}))^l},
$$
 and non-negative otherwise.
Note that when $l=2$, the AMEF is the Gaussian family, and since $\Gamma(\frac{1}{2})=\sqrt{\pi}$,  we recover the negative entropy condition $\eta=\sigma^2<\frac{1}{2\pi e}$; and conclude that $\sigma<\frac{1}{\sqrt{2\pi e}}$, as already claimed above.

Finally, we can extend the AMEF differential entropy formula to {\em location-scale AMEF distributions} with $\mu$ a {\em location parameter} and $\sigma$ a {\em dispersion parameter}.
Let $y=\mu+\sigma x$, and $p_l(x;\theta,\mu,\sigma)=\frac{1}{\sigma}p_l(\frac{x-\mu}{\sigma};\theta)$ (where $p_l(x;\theta)$ denotes the standard AMEF distribution).
We have $\dy=\sigma\dx$ and by making a change of variable in the integral of Eq.~\ref{eq:H} (see Appendix), it follows that $H(Y)=H(X)+\log\sigma$ (thus always independent of the location parameter).

\begin{lemma}
The differential entropy of a location-scale absolute monomial exponential family of degree $l$ and location parameter $\mu$ and dispersion parameter $\sigma>0$ is available in closed-form as:
\begin{eqnarray}
H_{l,\mu,\sigma}^\theta(\theta) &=& a_l - \frac{1}{l}\log (-\theta) + \log\sigma,\\
H_{l,\mu,\sigma}^\eta(\eta) &=& b_l + \frac{1}{l}\log \eta  + \log\sigma,
\end{eqnarray}
where $a_l=\log 2+\log \Gamma\left(\frac{1}{l}\right)-\log l + \frac{1}{l}$ and $b_l=\log \frac{2\Gamma(\frac{1}{l})(el)^{\frac{1}{l}}}{l}$.
\end{lemma}

Note that scaling an AMEF amounts to scale its natural parameter since $\theta\frac{|x|^l}{\sigma^l}=\theta_\sigma |x|^l$ 
with $\theta_\sigma=\frac{\theta}{\sigma^l}$
(see Eq.~\ref{eq:amefpdf}).
When the support is restricted to $\calX=[0,\infty)$ instead of $\bbR$, we subtract the $\log 2$ from $F$ and $H$ formula.
For example, this is useful when considering mixtures of Rayleigh distributions instead of Gaussian distributions.

\section{MaxEnt upper bounds for GMMs\label{sec:mm}}

In order to apply the MaxEnt upper bounds $U_l$ for a GMM with probability density function  $m(x)=\sum_{c=1}^k w_c p(x;\mu_c,\sigma_c)$, we need to compute its
absolute raw moment  and plug this value into formula Eq.~\ref{eq:ul}.
By linearity of the expectation operator, we have:
$$
E_{m(x)}\left[|X|^r\right]= \sum_{i=1}^k w_i E_{p(x;\mu_i,\sigma_i)}\left[|X|^r\right].
$$
The raw geometric moments and absolute raw geometric moments for a {\em centered} Gaussian distribution (ie. $\mu=0$) are reported in~\cite{normalmoments-2012}:
Closed-form formula are reported for the (absolute) moments for real-valued $r>-1$ using the Kummer's confluent hypergeometric functions~\cite{normalmoments-2012}.

For $l=2$, we can thus recover the well-known {\em MaxEnt Variance GMM upper bound}~\cite{KL-M-2016}:
\begin{equation}\label{eq:MEUBGMM}
H(X)\leq U_2=\frac{1}{2}\log\left(2\pi e \sum_{i=1}^k w_i((\mu_i-\bar\mu)^2+\sigma_i^2)\right), 
\end{equation}
with $\bar\mu=\sum_{i=1}^k w_i\mu_i$.

Surprisingly, we did not find the general formula for the raw absolute moments of a non-centered Gaussian.
We carried out the calculations reported in the Appendix.
Fortunately, the raw absolute moments of a Gaussian
 admit  closed-form formula expressed equivalently either using the Cumulative Distribution Function (CDF) $\Phi(x)=\int_{-\infty} \frac{1}{\sqrt{2\pi}}e^{-\frac{x^2}{2}} \dx$, the error function $\erf(x)$ or the complementary error function $\erfc(x)=1-\erf(x)$.
Those basic CDF, erf and erfc functions are related to each other by the following identities:
$$
\Phi(x)=\frac{1}{2}\left(1+\erf\left(\frac{x}{\sqrt{2}}\right)\right)=\frac{1}{2}\erfc\left(-\frac{x}{\sqrt{2}}\right).
$$
Based on our calculations, we state the series of MaxEnt upper bounds for the differential entropy of a GMM $X\sim m(x)=\sum_{c=1}^k w_cp(x;\mu_c,\sigma_c)$ in the following corollary of Theorem~\ref{theo1}:

\begin{corollary}
The differential  entropy $H(X)$ of a Gaussian mixture model $X\sim m(x)=\sum_{c=1}^k w_c p(x;\mu_c,\sigma_c)$ is upper bounded by:
$$
H(X)\leq U_l(X)= b_l+\frac{1}{l}\log A_l(X), \quad\forall l\in\bbN,
$$
where $b_l=\log 2 +\log \Gamma\left(\frac{1}{l}\right)-\log l + \frac{1}{l}(1+\log l)$ and
\begin{equation*}
A_l(X)=\left\{
\begin{array}{ll}
\sum_{c=1}^k w_c \sum_{i=0}^{\floor{\frac{l}{2}}}  \binom{l}{2i}  \mu_c^{l-2i}\sigma_c^{2i} 2^i \frac{\Gamma(\frac{1+2i}{2})}{\sqrt{\pi}} & \mbox{for even $l$},\\
\sum_{c=1}^k w_c \sum_{i=0}^l \binom{n}{i} \mu_c^{l-i}\sigma_c^i \left(I_i\left(-\frac{\mu_c}{\sigma_c}\right) - (-1)^i I_i\left(\frac{\mu_c}{\sigma_c}\right)  \right) & \mbox{for odd $l$}.
\end{array}
\right.,
\end{equation*}
where 
\begin{eqnarray*}
I_i(a)&=&\frac{1}{\sqrt{2\pi}}\int_a^{+\infty} x^i \exp\left(-\frac{1}{2}x^2\right)\dx,\\
&=& \frac{1}{\sqrt{2\pi}} \left( a^{i-1}\exp\left(-\frac{1}{2}a^2\right)  \right)
+(i-1)I_{i-2}(a),
\end{eqnarray*}
with the terminal recursion cases:
\begin{eqnarray*}
I_0(a) &=&  1-\Phi(a) = \frac{1}{2}\left(1-\erf\left(\frac{a}{\sqrt{2}}\right)\right)=\frac{1}{2}\erfc\left(\frac{a}{\sqrt{2}}\right),\\ 
I_1(a) &=& \frac{1}{\sqrt{2\pi}} \exp\left(-\frac{1}{2}a^2\right).
\end{eqnarray*}
\end{corollary}

In particular, the first two MaxEnt upper bounds (corresponding to the Laplacian and Gaussian MaxEnt  distributions, respectively) are given as:

\begin{corollary}[Laplacian maximum entropy upper bound]
The differential entropy of a Gaussian mixture model $X\sim \sum_{c=1}^k w_c p(x;\mu_c,\sigma_c)$ is upper bounded by:
$$
H(X)\leq U_1(X)=\log\left(2e\left(\sum_{c=1}^k w_c \left(\mu_c\left(1-2\Phi\left(-\frac{\mu_c}{\sigma_c}\right)\right) +  \sigma_c \sqrt{\frac{2}{\pi}}\exp\left(-\frac{1}{2}\left(\frac{\mu_c}{\sigma_c}\right)^2\right)\right)\right)\right).
$$
\end{corollary}

\begin{corollary}[Gaussian maximum entropy upper bound]
The differential entropy of a GMM $X\sim \sum_{c=1}^k w_c p(x;\mu_c,\sigma_c)$ is upper bounded by:
$$
H(X)\leq U_2(X)=\frac{1}{2}\log\left(2\pi e \sum_{c=1}^k w_c((\mu_c-\bar\mu)^2+\sigma_c^2)\right), 
$$  
with $\bar\mu=\sum_{c=1}^k w_c\mu_c$. 
\end{corollary}

Thus we can bound the differential entropy by $H(X)\leq\min (U_1(X),U_2(X))$, and for our series of upper bounds by:
$$
H(X)\leq\min_{i\in\bbN} U_i(X).
$$

Since the differential entropy does not change by changing the location parameter, we may consider without loss of generality that the GMM is centered to zero (that is, its expectation $E[X]$ is zero). If not, we simply translate the GMM by setting the component means to $\mu_i'=\mu_i-\bar\mu$ so that the expectation of the GMM matches the expectation of the AMEF. This alignment of the GMM to the AMEF  preserves the MaxEnt upper bounds.

In general, we may shift the GMM $X$ by $\delta\in\bbR$ by setting $\mu_i(\delta)=\mu_i-\delta$.
Let $X_\delta$ denotes this shifted GMM, $ A_l(\delta)=E[|X_\delta|^l]$ and $U_l(\delta)=U_l(X_\delta)$.
We can further refine the MEUBs  by minimizing the MaxEnt upper bounds:

$$
U_l(\delta)=  b_l+\frac{1}{l}\log A_l(\delta),
$$
where $b_l=\log 2 +\log \Gamma\left(\frac{1}{l}\right)-\log l + \frac{1}{l}(1+\log l)$.
 
When $l=2$, the optimal shift is obtained for $\delta=\bar\mu$.
However, when $l=1$, the optimization problem is non-trivial and requires numerical optimization procedures.
(In the remainder, we consider $\delta=\bar\mu$ when carrying experiments.)

\section{Experiments and tightness of the bounds\label{sec:exp}}

\subsection{Laplacian versus Gaussian MaxEnt upper bounds}

First, we consider the following experiment repeated $t=1000$ times:
We draw of a GMM $X\sim m$ with two components with $\mu_i,\sigma_i \simiid U(0,1)$ and $w_i\simiid U(0,1)$ chosen as uniform weights renormalized to $1$,
 we recenter the GMM so that $\bar\mu'=0$ (setting $\delta=\bar\mu$), and compute the stochastic approximation $\hat H$ of $H(X)$ (for $s=10^6$ samples), and the first order and second order maximum entropy upper bounds $H_1^\eta(E_m[|X|])$ and $H_2^\eta(E[X^2])$. We report  average approximations
 $\left|\frac{\hat H_i^\eta-\hat H}{\hat H}\right|$, and the percentage of times MEUB $U_1(X)<U_2(X)$:
Gaussian MEUB is on average $40\%$ above $\hat H$ and Laplacian MEUB is on average $10\%$ above $\hat H$.
Laplacian MEUB bound beats the Gaussian MEUB $32.9\%$ on average.
Thus we recommend practitioners to upper bound the differential entropy of a GMM $X$ by $H(X)\leq \min(U_1(X),U_2(X))$.

\subsection{Series of MaxEnt upper bounds}

A question one may ponder is whether all MEUBs $U_l(X)$ are useful of not?
We performed an experiment by drawing at random GMMs $X$ with $k=2$ components and checking among the first $n$ bounds $U_1(X), \ldots, U_n(X)$.
We found experimentally that most of the time the bounds $U_1(X)$ and $U_2(X)$ suffices, but sometimes the tightest bound could be $U_n(X)$.
This is the case when one component is almost a Dirac ($\sigma=o(1)$) while the other component has significant standard deviation, and the two Gaussian components far apart. 
For example, let us choose $X\sim m(x)=\frac{1}{2}p(x,-\frac{1}{2},10^{-5})+\frac{1}{2}p(x,\frac{1}{2},10^{-1})$.
Then $U_{i+1}(X)\leq U_(X)$ for $i\leq 37$.
We get NaN numerical errors when computing $U_{38}$ using the closed-form formula.

We shall make more precise those arguments in the following section.

\subsection{Tightness analysis of MaxEnt upper bounds}

First, let us show that bound $U_1$ (the Laplacian MEUB) may be better than $U_2$ (the Gaussian MEUB).
To derive analytic conditions, we consider the {\em restricted case} of zero-centered GMMs~\cite{zeroGMM-2002}.
We have  
 $A_1(X)=\sqrt{\frac{2}{\pi}}\sum_{i=1}^k w_i\sigma_i=\sqrt{\frac{2}{\pi}}\bar\sigma$, and therefore
get the upper bound $U_1$ on the differential entropy of the mixture as:

\begin{equation}
H(X) \leq U_1(X)=1+\log 2\sqrt{\frac{2}{\pi}}\left(\sum_{i=1}^k w_i\sigma_i\right).
\end{equation}

This bound is strictly better than the traditional Gaussian bound~\cite{H-GMMiso-2016}:
$$
U_2(X)=\log \sqrt{2\pi e}\sqrt{\sum_{i=1}^k w_i\sigma_i^2},
$$ 
provided that $U_1(X)< U_2(X)$. 
Note that when $l=2$ and $k=1$, the $U_2$ bound matches precisely the entropy of the single-component Gaussian mixture.
Let $\overbar{\sigma}_1$   be the {\em arithmetic weighted mean} and  
$\overbar{\sigma}_2=\sqrt{\sum_{i=1}^k w_i\sigma_i^2}$ be the {\em quadratic mean} of the weighted standard deviations, respectively.
Then $U_1(X) < U_2(X)$ if and only if:
\begin{equation}
\log 2e\sqrt{\frac{2}{\pi}}\bar{\sigma_1} \leq \log \sqrt{2\pi e}\overbar{\sigma_{2}}.
\end{equation}
That is, we need to have $\frac{\overbar{\sigma}_1}{\overbar{\sigma}_{2}} \leq \frac{\pi}{2\sqrt{e}} \approx 0.9527$.
Observe that the {\em weighted quadratic mean} dominates\footnote{In general, we denote for a strictly increasing function $f(x)$ the quasi-arithmetic  weighted mean by  
$\overbar{\sigma}_{f(x)}=f^{(-1)}\left(\sum_{i=1}^k w_i f(\sigma_i) \right)$, see~\cite{GQA-2015}.} the {\em weighted arithmetic mean}, and 
therefore $\frac{\overbar{\sigma}_1}{\overbar{\sigma}_{2}}\leq 1$.
Equality of arithmetic/quadratic means only happens when all the $\sigma_i$'s coincide (since we have zero-centered GMMs, that means that the GMM collapses to a Gaussian).
To summarize our illustrating example, bound $U_1$ may be better or worse than $U_2$ depending on the set of $\sigma_i$'s.
For the degenerate case $k=1$ (single component $X=N(0,\sigma)$), the condition of $U_1<U_2$ ($U_1$ tighter than $U_2$) writes as $\sigma>\frac{2\sqrt{e}}{\pi}$  (that is, $\sigma>1.0496$).

Now, for $l\in\bbN$, we built a MaxEnt upper bound on the differential entropy of a GMM $X\sim m(x)$.
How does this infinite sequence of bounds $U_1, U_2, \ldots, U_{q}, \ldots$ relate to each others?
For a prescribed value $A$, $H_l^\eta(A)$ decreases as $l$ increases, but the absolute raw moment $A_l(X)$ also varies. 
Does there always exist a GMM $X$ so that there exists $l'>l$ such that $H_{l'}^\eta(A_{l'}(X))<H_l^\eta(A_l(X))$, or not?

For general GMMs, we explained in the former section that by taking a GMM with two components with one component almost a Dirac, we could establish experimentally that all bounds could yield the tightest one.
Here, to answer negatively this question when considering the family of zero-centered Gaussian mixtures, 
we shall consider {\em even} integers $l$ and $l'=l+2$.
Let $N\sim N(\mu=0,\sigma)$. Then the geometric raw moments coincide with the central geometric moments, and by the linearity of the expectation operator, we have~\cite{normalmoments-2012}:
\begin{equation}
E_{X}[X^l]= \underbrace{2^{\frac{l}{2}}\frac{\Gamma(\frac{1+l}{2})}{\sqrt{\pi}}}_{z_l}  \left(\sum_{i=1}^k w_i \sigma_i^l \right) =A_l(X).
\label{eq:gmm0}
\end{equation}

Then we have the following MaxEnt upper bound $U_l$: 
\begin{lemma}[Zero-centered GMMs]
The differential entropy of a zero-centered GMM $X\sim \sum_{c=1}^k w_c p(x;0,\sigma_c)$ is upper bounded by:
\begin{equation}\label{eq:ugmm0}
H(X)\leq H_l^\eta(A_l(X)) = b_l+\frac{1}{l}\log z_l +\log \bar{\sigma}_l,
\end{equation}
where $\bar{\sigma}_l$ is the {\em $l$-th power mean}: 
\begin{equation}
\bar{\sigma}_l=\left(\sum_{i=1}^k w_i \sigma_i^l \right)^{\frac{1}{l}}.
\end{equation}
\end{lemma}
When $l\rightarrow \infty$, we have $\bar{\sigma}_{l}\rightarrow \max_i \sigma_i$.
Thus for a tighter bound $U_{l+2}<U_l$, we need to find a zero-centered GMM so that:

$$
\log \left(\frac{ \bar{\sigma}_{l+2}}{\bar{\sigma}_l}\right)\leq \underbrace{b_l+\frac{1}{l}\log z_l - b_{l+2} -\frac{1}{l+2}\log z_{l+2}}_{\Delta_l}. 
$$

Since the   $(l+2)$-power mean dominates\footnote{A mean $\overbar{\sigma}_{g(x)}$ dominates another mean $\overbar{\sigma}_{f(x)}$ (that is, $\overbar{\sigma}_{g(x)}\geq \overbar{\sigma}_{f(x)}$) when $g(x)\geq f(x)$.} the $l$-power mean (ie., $\bar{\sigma}_{l+2}>\bar{\sigma}_l$), we have $\frac{ \bar{\sigma}_{l+2}}{\bar{\sigma}_l}\geq 1$
 and therefore 
$\log \frac{ \bar{\sigma}_{l+2}}{\bar{\sigma}_l}\geq 0$.
It turns out that $\Delta_l<0$ when $l>2$ with $\lim_{l\rightarrow\infty} \Delta_l=0$.
So we conclude that only $U_1$ and $U_2$ are necessary for zero-centered GMMs (and we define the MEUB as $\min(U_1,U_2)$).

When considering arbitrary GMMs, the situation is analytically more complex to decide.

Last but not least, whether bound $U_l$ proves useful or not depends on the mixture family (eg., mixtures of Pareto distributions~\cite{Pareto-2013}).
A Pareto distribution has density $\frac{\alpha}{x^{\alpha+1}}$ for $x>0$ and $\alpha$ the shape parameter.
The raw moments of a Pareto distribution is $A_l=\frac{\alpha}{\alpha-l}$ for $\alpha>l$ and $\infty$ otherwise.

\section{Conclusion\label{sec:concl}}
We considered the novel parametric family of Absolute Monomial Exponential Families (AMEFs), and reported a closed-form differential entropy formula for these AMEFs.
We then considered a collection of Maximum Entropy Upper Bounds (MEUBs) for an arbitrary continuous random variable based on its raw geometric absolute moments (Theorem~\ref{theo1}), and show how to apply those generic bounds to the specific case of Gaussian Mixture Models (GMMs).
Interestingly, we showed that the Laplacian MaxEnt upper bound may potentially be tighter than the traditionally used Gaussian MaxEnt upper bound. 
Therefore, we recommend in practice to take the minimum of these Laplacian and Gaussian  MEUBs.
This new series of MaxEnt upper bounds proves useful in practice since the differential entropy of mixtures does not admit a closed-form formula~\cite{KLnotanalytic-2004}.
Besides, we report in the Appendix closed-form formula for calculating the  raw absolute moments of a univariate Gaussian Mixture Model.
The method can be extended to any location-scale univariate continuous distribution.

A Java\texttrademark{} source code for reproducible research with test experiments is available  at:\\
\begin{center}
\url{https://www.lix.polytechnique.fr/~nielsen/MEUB/} 
\end{center}

\appendix

\section{Differential entropy of a location-scale distribution\label{app:diffent}}

Let $p(x;\mu,\sigma)=\frac{1}{\sigma}p_0(\frac{x-\mu}{\sigma})$ denote the density of a {\em location-scale distribution} on the full support $\bbR$, where $\mu\in\bbR$ denotes the {\em location parameter} and $\sigma>0$ the {\em dispersion parameter}.
For example, a normal distribution has location parameter its mean and dispersion parameter its standard deviation.
Let us prove that the entropy $H(X)$ is $H(X_0)+\log\sigma$ with $X\sim p(x;\mu,\sigma)$ and $X_0\sim p_0(x)$, a quantity always independent of the location parameter $\mu$.
We shall make use of a change of variable $y=\frac{x-\mu}{\sigma}$  (with $\dy=\frac{\dy}{\sigma}$) in the integral to get:
\begin{eqnarray}
H(X)&=&\int_{x=-\infty}^{+\infty} -\frac{1}{\sigma}p_0\left(\frac{x-\mu}{\sigma}\right)\left(\log \frac{1}{\sigma}p_0\left(\frac{x-\mu}{\sigma}\right)\right) \dx,\\
&=& \int_{y=-\infty}^{+\infty} - p_0(y)(\log p_0(y) -\log\sigma),\\
&=& H(X_0)+\log\sigma .
\end{eqnarray}

\section{Raw absolute moments of a non-centered normal distribution\label{app:moment}}

Let $X \sim N(\mu,\sigma)$ be a normal random variable of mean $\mu\in\bbR$ and standard deviation $\sigma>0$. 
Let us express the density of the normal distribution as a location-scale density: $p(x;\mu,\sigma)=\frac{1}{\sigma}p_0(\frac{x-\mu}{\sigma})$,
where $p_0(x)$ denotes the density of the standard normal distribution $X_0$:
$$
p_0(x)=\frac{1}{\sqrt{2\pi}} \exp\left(-\frac{1}{2}x^2\right).
$$

Define the raw (uncentered) $l$-th absolute moment $E[|X|^l]$ for a continuous univariate location-scale family with standard density $p_0$:
$$
A_l = E[|X|^l] = \int_{-\infty}^{+\infty} |x|^l \frac{1}{\sigma} p_0\left(\frac{x-\mu}{\sigma}\right) \dx .
$$

We first consider the calculation of  $A_l=E[|X|^l]=E[X^l]$ for even integer $l$, and then proceed with the computation of $E[|X|^l]$ for odd $l$.

\subsection{Raw even (absolute) moments}
The raw absolute geometric moment amounts to the raw geometric moment for even integer $l$: $a_l=E[|X|^l]=E[X^l]$.
 It follows after a change of variable $y=\frac{x-\mu}{\sigma}$ (so that $x=\sigma y+\mu$) with $\dy=\frac{\dx}{\sigma}$ (and $\dx=\sigma \dy$) that:
$$
A_l=  \int_{-\infty}^{+\infty} (\sigma y+\mu)^l p_0(y) \dy.
$$

Performing the {\em binomial expansion}  $(\sigma y+\mu)^l=\sum_{i=0}^l \binom{l}{i}(\sigma y)^i \mu^{l-i} $, we get:
$$
A_l=  \sum_{i=0}^l \binom{l}{i} \mu^{l-i} \sigma^i \int_{-\infty}^{+\infty} y^i p_0(y) \dy,
$$
where $\int_{-\infty}^{+\infty} y^i p_0(y) \dy$ is the $i$-th raw moment of the standard normal distribution $X_0$:
$$
E[|X_0|^l]=\left\{
\begin{array}{ll}
(l-1)!!\sigma^l & \mbox{even $l$},\\
\sqrt{\frac{2}{\pi}} 2^{\frac{l-1}{2}} \left(\frac{l-1}{2}\right)! \sigma^l & \mbox{odd $l$}.
\end{array}
\right.,
$$
with $n!!$ the double factorial: $n!!=\sqrt{\frac{2^{n+1}}{\pi}}\Gamma(\frac{n}{2}+1)$. 

We end-up with the following raw moment {\em direct formula} for an even integer $l$:

$$
A_l = \sum_{i=0}^{\floor{\frac{l}{2}}}  \binom{l}{2i}  \mu^{l-2i}\sigma^{2i} 2^i \frac{\Gamma(\frac{1+2i}{2})}{\sqrt{\pi}}.
$$

In particular, we recover the second (absolute) moment:
$$
E\left[|X|^2\right] = \mu^2+\sigma^2.  
$$

\subsection{Raw absolute moments of odd order}

We get rid of the cumbersome absolute value by splitting the integral onto the positive and negative support as follows:

$$
A_l = \int_{-\infty}^{0} -x^l \frac{1}{\sigma} p_0\left(\frac{x-\mu}{\sigma}\right) \dx + 
\int_{0}^{+\infty} x^l \frac{1}{\sigma} p_0\left(\frac{x-\mu}{\sigma}\right) \dx 
$$ 

Consider the change of variable $y=\frac{x-\mu}{\sigma}$.
We get:

$$
A_l =  {\int_{-\infty}^{\frac{-\mu}{\sigma}} -(\sigma y+\mu)^l   p_0(y) \dy}  + 
 {\int_{\frac{-\mu}{\sigma}}^{+\infty} (\sigma y+\mu)^l  p_0(y) \dy} . 
$$
 
By performing binomial expansion and sliding the integral inside the binomial sum, we get:
$$
A_l=  \sum_{i=0}^l \binom{n}{i} \mu^{l-i}\sigma_i \left(  I_i\left(-\frac{\mu}{\sigma}\right) - J_i\left(-\frac{\mu}{\sigma}\right)  \right),
$$
with
\begin{eqnarray}
I_i(a)&=& \frac{1}{\sqrt{2\pi}}\int_a^{+\infty} x^i \exp\left(-\frac{1}{2}x^2\right)\dx,\\
J_i(b)&=&  \frac{1}{\sqrt{2\pi}}\int_{-\infty}^b x^i \exp\left(-\frac{1}{2}x^2\right)\dx.
\end{eqnarray}

By a change of variable $y=-x$ (with $\dy=-\dx$), we find that:

$$
J_i(b)=(-1)^i I_i(-b).
$$

Thus it is enough to consider the computation of $I_i$ for all non-negative integers $i\geq 0$, and get the raw absolute moment as:
$$
A_l=  \sum_{i=0}^l \binom{n}{i} \mu^{l-i}\sigma^i \left(  I_i\left(-\frac{\mu}{\sigma}\right) - (-1)^i I_i\left(\frac{\mu}{\sigma}\right)  \right),
$$

Consider the {\em integration by parts}\footnote{Recall that $\int_a^b u(x)v'(x)\dx=[u(x)v(x)]_a^b-\int_a^b u'(x)v(x)\dx$.} 
for calculating integral:
$$
I_i(a)=\frac{1}{\sqrt{2\pi}}\int_a^\infty x^i \exp\left( -\frac{1}{2}x^2 \right)\dx,\quad i\in\bbN
$$
with $v'(x)=x\exp\left(-\frac{1}{2}x^2\right)$  (and antiderivative $v(x)=-\exp\left(-\frac{1}{2}x^2\right)+c$, where $c$ is a constant) 
and $u(x)=\frac{1}{\sqrt{2\pi}} x^{i-1}$ (and derivative $u'(x)=\frac{1}{\sqrt{2\pi}}(i-1)x^{i-2}$):

$$
I_i(a)= \frac{1}{\sqrt{2\pi}} a^{i-1}\exp\left(-\frac{1}{2}a^2\right)
- (i-1) \frac{1}{\sqrt{2\pi}}  \int_a^\infty x^{i-2} \exp\left(-\frac{1}{2}x^2\right)\dx .
$$

Thus we end up with the following recursive formula: 
$$I_i(a)= \frac{1}{\sqrt{2\pi}} \left( a^{i-1}\exp\left(-\frac{1}{2}a^2\right)  \right)
+(i-1)I_{i-2}(a),
$$

with the terminal recursion cases:

\begin{eqnarray}
I_0(a) &=&  1-\Phi(a),\\ 
I_1(a) &=& \frac{1}{\sqrt{2\pi}} \exp\left(-\frac{1}{2}a^2\right).
\end{eqnarray}

Equivalent expressions may be obtained using the error function or the complementary error function by using the following identities:
$$
\Phi(x)=\frac{1}{2}\left(1+\erf\left(\frac{x}{\sqrt{2}}\right)\right)=\frac{1}{2}\erfc\left(-\frac{x}{\sqrt{2}}\right).
$$

\subsection{Numerical robustness}
We  checked experimentally the numerical robustness of those formula by comparing the exact moment formula with the Monte-Carlo estimated ones.
In general, the moment $E_X[g(X)]=\int g(x)p(x;\mu,\sigma)\dx$ can be approximated stochastically using Monte-Carlo sampling as:
$$
\frac{1}{s}\sum_i g(x_i),
$$
with $x_1,\ldots, x_s\simiid X$.
MC estimators are consistent (ie., tend to the true value when $s\rightarrow\infty$).
We can also discretize the moment integral using various quadratic rules to approximate the moment values.

The table below reports the first ten (10) raw absolute moments $E[|X|^l]$ of a Gaussian random variable $X\sim N(\mu=1,\sigma=2)$:
 \begin{center}
{\small
\begin{tabular}{llll}
Order $l$ & MC estimation ($s=10^8$) & formula & error in percent \%  \\ \hline
0       & 1.0   &   0.9999999999811017     & 1.8898327347471877E-9\\
1       & 1.7911208288723894      & 1.7911862296052241    &  0.003651385868581992\\
2       & 5.001255187253235       & 4.999999999721595     &  0.025097450232866934\\
3       & 17.65128588616714       & 17.652375756639124    &  0.006174453685761998\\
4       & 73.02031705881072       & 72.99999999342585     &  0.027823852597777112\\
5       & 339.9322718149212       & 339.96501890043305    &  0.009633414720237885\\
6       & 1740.1120429182856      & 1740.9999997924972    &  0.051028718399213205\\
7       & 9659.309994586874       & 9649.665608394966     &  0.099845498253121\\
8       & 57314.11839221297       & 57232.999991840465    &  0.14153301603174934\\
9       & 359089.90092220495      & 360173.10941100196    &  0.30165384379096866\\
10      & 2400947.7149177645      & 2389140.9996155496     & 0.4917522871846192\\
\end{tabular}
}
\end{center}

The error percentage is defined as $\frac{100|\mathrm{MC}-\mathrm{formula}|}{\mathrm{MC}}$.
Note that although the formula of the raw absolute moment is exact, we use a fast approximation of the gamma function (see the numerical receipe in~\cite{nr-2007}).

We report below  the first fifteen (15) raw absolute moments $E[|X|^l]$ of a Gaussian mixture random variable $X\sim m(x)=\frac{1}{2}p(x;-1,1)+\frac{1}{2}p(x;1,1)$:

\begin{center}
{\small
\begin{tabular}{llll}
Order $l$ & MC estimation ($s=10^8$) & formula & error in percent \%  \\ \hline
0 & 1.0 & 1.0 & 0.0 \\
1 & 1.0230962043303955 & 1.022982026195441 & 0.011160058503905857\\
2 & 1.5868516647073658 & 1.5870832593136879 & 0.01459459705483931\\
3 & 3.06462511302367 & 3.0653736036966404 & 0.024423563906384612\\
4 & 6.868131662942202 & 6.869191248884848 & 0.015427571785835445\\
5 & 17.223233174871346 & 17.231439294893978 & 0.047645642019200164\\
6 & 47.33241385186923 & 47.34851815902828 & 0.034023845074654065\\
7 & 140.3877346937668 & 140.44292724406705 & 0.03931436775488671\\
8 & 444.72048147704123 & 444.9348623596309 & 0.04820575878980772\\
9 & 1493.5450728510505 & 1493.4616390491788 & 0.005586292867106997\\
10 & 5273.891071845998 & 5277.575633794758 & 0.06986420270280147\\
11 & 19493.58060503037 & 19533.828198789193 & 0.20646588522807466\\
12 & 75646.96071848647 & 75407.19940214818 & 0.3169477187993617\\
13 & 302766.1898278726 & 302530.01691011444 & 0.0780050499999491\\
14 & 1260210.5985202221 & 1257605.9222275713 & 0.20668579487502786\\
15 & 5354365.384185081 & 5402772.331681476 & 0.904065076308983
\end{tabular}
}
\end{center}

\subsection{Raw sufficient statistic moments of exponential families}

For univariate mixtures of natural EFs with a polynomial sufficient statistic $t(X)$, we may easily calculate moments using the Moment Generating Function (MGF)~\cite{EF-1986}:

\begin{equation}
M(t(u))=E[\exp(\inner{u}{X})]=\exp(F(\theta+u)-F(\theta)).
\end{equation}

Thus for uni-order EFs, the geometric moments are given by the higher-order derivatives $E_X[t(X)^l]=M^{(l)}(0)$.
For uni-order exponential families, it follows that $E[t(X)]=F'(\theta)=\eta$, and $V[t(X)]=F''(\theta)>0$ (since $F$ is strictly convex).
It follows that EFs have always  all finite order moments expressed using the higher-order derivatives of the MGF.
Thus we can always explicitly calculate the geometric moments of the sufficient statistic $t(X)$ from the MGF provided that the log-normalizer $F(\theta)$ is available in closed-form. For example, we may consider mixtures of Rayleigh distributions (with $t(x)=x^2$, see~\cite{EF-1986}) instead of GMMs, and get closed-form MaxEnt upper bounds.
The geometric raw moments of a Rayleigh mixture $X$ is $A_l(X)=2^{\frac{l}{2}}\Gamma(1+\frac{l}{2})\sum_{i=1}^k \sigma_i^l$.

\section*{Symbolic integration using a computer algebra system\label{app:maxima}}

The examples below show how definite integration is performed using the Computer Algebra System (CAS) {\tt Maxima}\footnote{\url{http://maxima.sourceforge.net/}}:

{
\begin{verbatim}
/* Example of a density that has infinite
 Shannon entropy */
p(x) := log(2)/(x*log(x)**2);
/* check it integrates to 1 */
integrate(p(x),x,2,inf);
/* check integral diverges */
integrate(-p(x)*log(p(x)),x,2,inf);
\end{verbatim}
}
To handle AMEFs, we need to enforce that $\theta<0$.
For example, to compute the log-normalizer of an AMEF of order $l=5$ in {\tt Maxima}, we may use the following script: 
{ 
\begin{verbatim}
assume (theta<0);
F(theta) := integrate(exp(theta*abs(x)^5),x,-inf,inf);
integrate(exp(theta*abs(x)^5-F(theta)),x,-inf,inf);
\end{verbatim}
}
 
To program a binomial expansion in  {\tt Maxima}, we write the following recursive function:
{ 
\begin{verbatim}
binomialExpansion(i,p,q) := if i = 1 then p+q
else expand((p+q)*binomialExpansion(i-1,p,q)) ;

expand(binomialExpansion(10,x,y));
factor(%);
\end{verbatim}
}

We get at the console the following output:

\noindent\includegraphics[width=\textwidth]{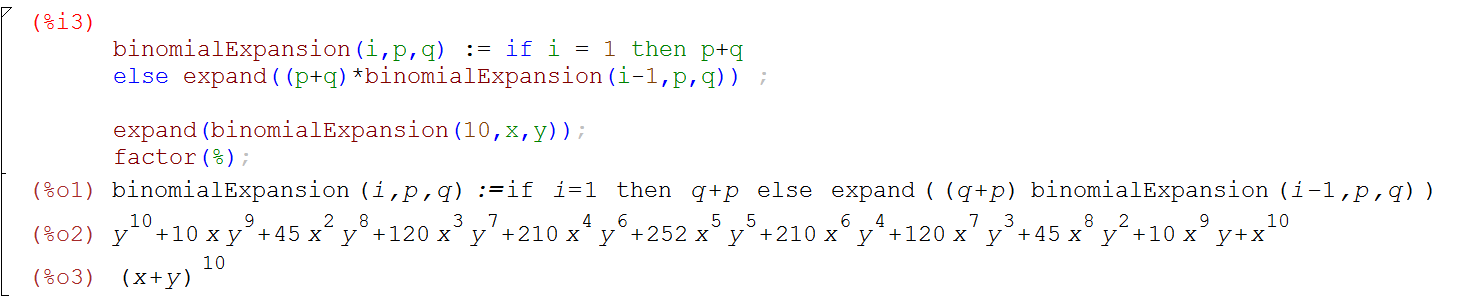}

Thus to obtain direct formulae for the raw absolute moments, we may use the following symbolic program in {\tt Maxima}:
\begin{verbatim}
binomialExpansion(i,p,q) := if i = 1 then p+q
else expand((p+q)*binomialExpansion(i-1,p,q)) ;

p0(y) := exp(-y^2/2)/sqrt(2*pi);

absMoment(mu,sigma,l) := 
ratsimp(integrate(factor(expand(binomialExpansion(l,mu,y*sigma)))*p0(y),y,-mu/sigma,inf)
-integrate(factor(expand(binomialExpansion(l,mu,y*sigma)))*p0(y),y,-inf,-mu/sigma));

assume(sigma>0);
absMoment(mu,sigma,1);
absMoment(mu,sigma,3);
\end{verbatim}

We get at the console the following output:
 
\noindent\includegraphics[width=\textwidth]{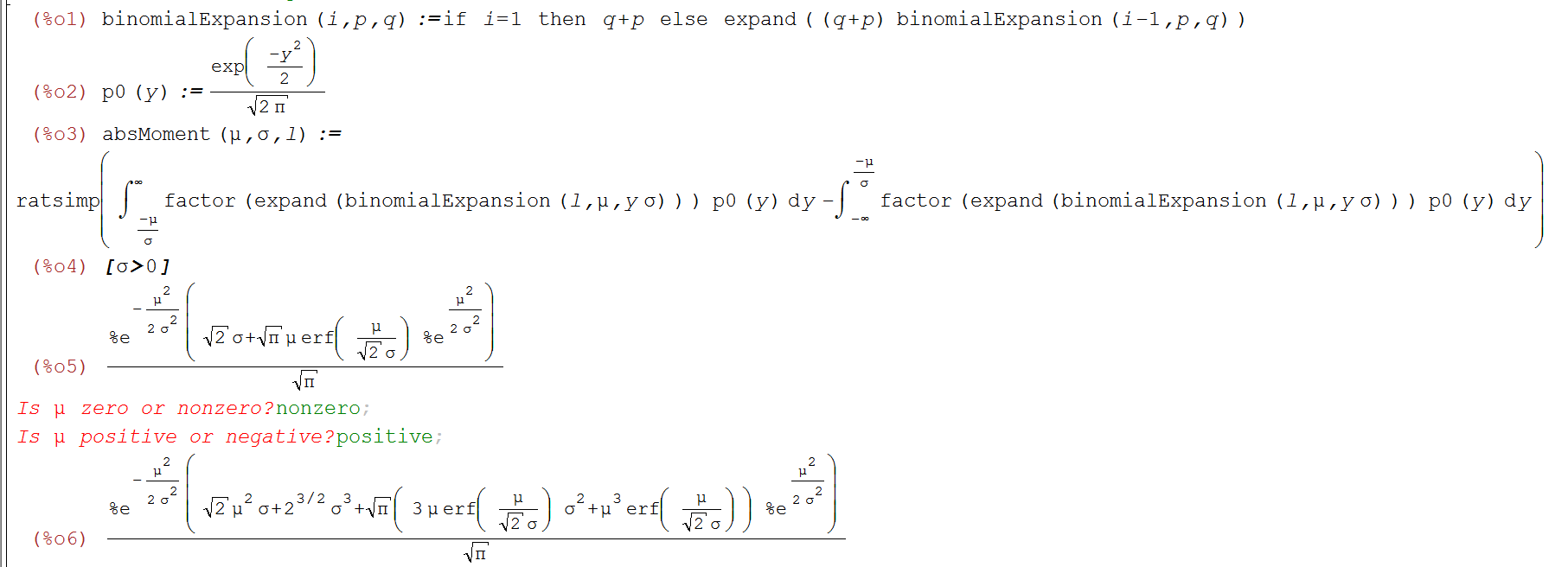}


\end{document}